\documentclass[rapid]{JFM-FLM_Au}
\usepackage[english]{babel}
\usepackage{bm}
\usepackage[percent]{overpic}
\usepackage{xfp}
\usepackage{csquotes}
\usepackage{placeins}
\usepackage[switch]{lineno}
\usepackage{sansmath}

\hypersetup{
  colorlinks = true,
  citecolor  = blue,
  linkcolor  = blue,
  urlcolor   = blue
}

\lefttitle{Bagheri, Becker \& Schlatter}
\righttitle{Journal of Fluid Mechanics Rapids}

\title{Collapse of turbulence in optimised curved pipe flow}

\author{
  Eman Bagheri\aff{1},
  Stefan Becker\aff{1}
  \and
  Philipp Schlatter\aff{1}
}

\corresau{Eman Bagheri\, \email{eman.m.bagheri@fau.de}}

\affiliation{\aff{1}Institute of Fluid Mechanics (LSTM),
Friedrich--Alexander--Universit\"at (FAU) Erlangen--N\"urnberg,
91058 Erlangen, Germany}

\makeatletter
\renewcommand{\oddfooterflagdefns}[1]{}
\renewcommand{\evenfooterflagdefns}[1]{}
\renewcommand{\oddabsfooterflag}{}
\renewcommand{\evenabsfooterflag}{}

\providecommand{\section}[2][]{%
  \section*{#2}%
}

\makeatother

\begin{document}

\maketitle
% ------------------------------------------------------------------
% Abstract & keywords
% ------------------------------------------------------------------
\begin{abstract}
Turbulence-induced friction is a significant contributor to energy consumption in the fluid-transport and piping industries. Here, we describe a passive approach to suppress turbulence and reduce friction: we show that a local increase in streamwise flow curvature, combined with changing the circular cross section to an oval, relaminarises turbulent flow in curved pipes. We exemplify this effect in a \(180^{\circ}\) bend at \(Re_D=10,000\) and \(20,000\) (based on bulk velocity \(U_B\) and pipe diameter \(D\)), well above the limit for sustained turbulence in straight pipes and the linear stability limit in  \(180^{\circ}\) bends. Curvature inhibits streamwise Reynolds stresses, and cross-sectional modifications weaken the unstable secondary flow, together disrupting the near-wall regeneration cycle and collapsing turbulence. Simulations and experiments confirm that these geometric modifications suppress turbulence and reduce pressure loss by 53\% and 36\% compared to the baseline \(180^{\circ}\) bend and a fully developed straight pipe of equal length, respectively. The results establish a passive, mechanism-based route to relaminarisation in curved pipes with implications for energy-efficient control in other wall-bounded flows with curvature.
\end{abstract}


% ------------------------------------------------------------------
% Main text
% ------------------------------------------------------------------

\section{Introduction}
In most applications, a key objective is to reduce turbulence as much as possible because of its adverse effect on frictional losses. It has been demonstrated previously that by appropriately modifying the mean profile of a turbulent pipe flow, partial or full relaminarisation may be achieved \citep{Hof2010,Kuehnen2018,Scarselli2019}. In curved pipes, frictional losses are even more pronounced, given the thinner shear layers on the outer side of the bend, coupled with geometry-induced in-plane motion.  
Flow in curved pipes is integral to applications such as heat exchangers, biological systems (e.g., the aortic arch), and fluid transport networks, where turbulence-induced frictional loss remains a major contributor to global energy consumption \citep{Berger1983,Vashisth2008, Vester2016}. Despite more than a century of research, several open questions related to curved-pipe flows remain. From the seminal contributions of \cite{Dean1927, Dean1928}, it is well established that curvature induces a secondary flow, the so-called Dean vortices. Subsequently, \cite{White1929} and \cite{Taylor1929} studied turbulence and transition in curved pipes and discovered that curvature fundamentally changes both the onset and the dynamics of turbulence. Although the influence of curvature on transition and turbulence has been studied and partial curvature-induced flow relaminarisation has since been reported several times in the literature \citep{Sreenivasan1983,Noorani2013_IJHFF}, the underlying mechanisms that can suppress turbulence are not yet well understood. In this context, two important open questions remain: Can geometric features disrupt the turbulence regeneration cycle and restore laminar flow beyond the stability limit? If so, through what physical mechanisms is relaminarisation achieved? %We demonstrate that this is indeed possible through passive modifications of the curvature and cross-sectional profile, which suppress near-wall streaks and lead to relaminarisation.
%s\vspace{-0.5em}

As energy consumption and pumping power have long been of major concern, the focus of many earlier studies was the quantification of the friction factor and pressure loss. \cite{Ito1959, Ito1960} made fundamental contributions by providing widely used empirical correlations for the friction factor and the bend loss coefficient in smooth bent pipes as functions of Reynolds number and curvature. \cite{Sreenivasan1983} then provided experimental evidence that curvature can stabilise turbulence in helically coiled pipes and even induce relaminarisation at moderate Reynolds numbers. More recently, \cite{Kuehnen2015} confirmed Sreenivasan's conclusion that a fully turbulent flow may be relaminarised due to curvature effects. \cite{Cioncolini2006} studied a wide range of non-dimensional curvatures ($\gamma = D/(2R_c) \simeq 0.0027-0.145$) and Reynolds numbers ($\mathrm{Re_D}=U_BD/\nu\approx 10^{3}-6.3\times10^{4}$) and found that, outside of the transitional region, their data were generally in agreement with Ito’s correlations. Here, $R_c$ denotes the centreline's radius of curvature, $D$ is the pipe diameter, and $U_B$ represents the bulk velocity. However, for coils with curvature of $\gamma \simeq 0.01\text{–}0.03$, they reported a \enquote{depression} region in the friction–factor profile within which the curved–pipe friction factor is not necessarily higher than that of a straight pipe at the same Reynolds number (i.e.,\ \enquote{substraight} behaviour). Using direct numerical simulations (DNSs) at $\gamma = 0.01$, \cite{Noorani2013_IJHFF} also found a substraight regime at $\mathrm{Re_D}=5300$. \cite{Noorani2015_PoF} subsequently discovered a regime \enquote{sublaminar} near $\mathrm{Re_D} \approx 3400$ in long, mildly curved pipes ($\gamma = 0.01$), in which the mean drag was up to 8\% below the laminar correlation.  Furthermore, they observed that strong curvature ($\gamma = 0.1$) partially relaminarises the flow near the inner wall. %This is due to Pss and Prr curvature terms

Consistently, earlier DNS of helically coiled pipes by \cite{Huettl2000a} also showed that strong curvature ($\gamma = 0.1$) suppresses turbulence and can nearly relaminarise the flow. However, they also noted that the secondary flow can generate remarkably high local Reynolds shear stresses near the upper and lower walls, underscoring the impact of the secondary flow in sustaining turbulence. In transitional bent pipe flows, \cite{Rinaldi2019} showed that curvature eliminates the strong upstream front of puffs and slugs and pushes the production region toward the outer wall. Overall, the curvature-induced relaminarisations in the literature are reported at low Reynolds numbers ($\mathrm{Re_D}\approx 3\times 10^3$--\ $6\times 10^3$).~\cite{Canton2016} indicate that these low Reynolds number stabilising effects observed in moderate-curvature regimes are attributed to the linear stability threshold in curved pipes, rather than a turbulence suppression mechanism.  At higher $\mathrm{Re_D}$, only the inner wall region partially relaminarises while the overall friction factor remains higher than that of the straight pipe. 

Previous studies have established that turbulence in curved pipes is governed by the Reynolds number and curvature. The collective evidence shows that curvature introduces competing mechanisms that can promote or suppress turbulence. In this context, two aspects remain unexplored in the literature. Firstly, the evidence for relaminarisation in curved pipes at high Reynolds numbers well beyond the linear stability limit is absent to the best of our knowledge. Secondly, the effect of the cross-sectional profile on the Dean vortices and their influence on turbulence has been overlooked. The latter is crucial since a strong streamwise curvature can indirectly intensify turbulence by transferring energy from the primary flow to unstable Dean vortices. On the other hand, an appropriate cross-sectional shape modification can weaken the Dean vortices, thereby limiting their role in both the production and redistribution of turbulence.
\vspace{-4mm} 
\section{Motivation and methodology}
\begin{figure}[h]
  \centering
  \begin{overpic}[width=0.75\linewidth,trim={0 10 0 1},clip]{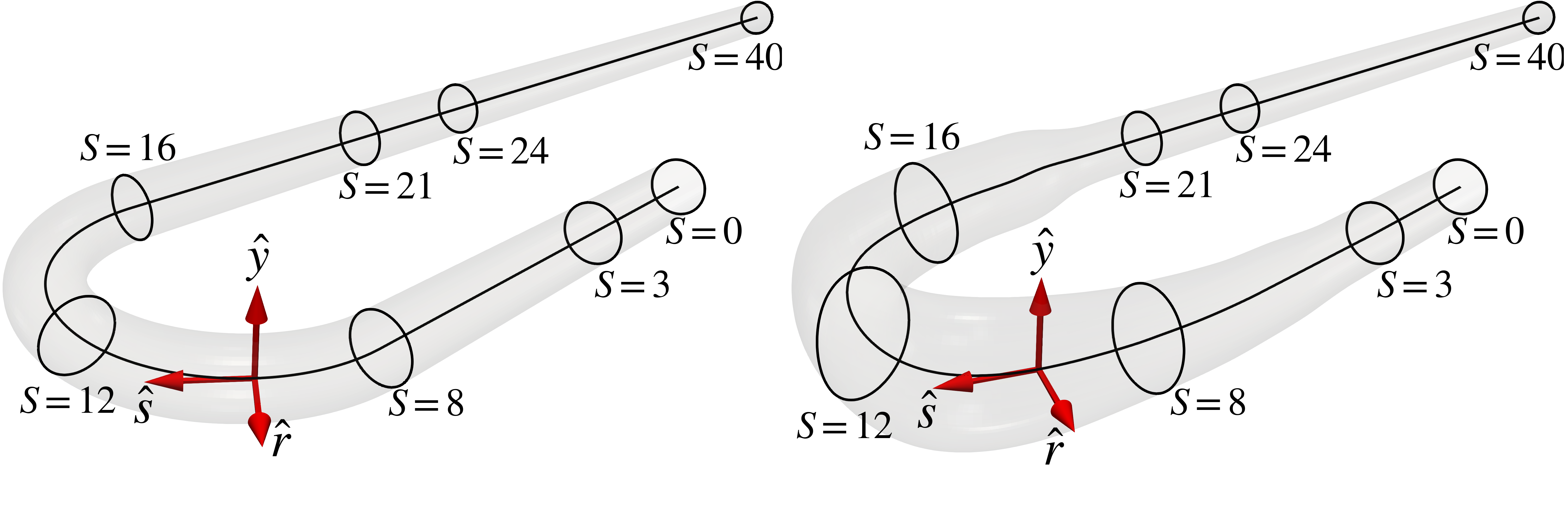}
    \put(21,1){\textbf{(a)}}
    \put(70,1){\textbf{(b)}}
  \end{overpic}
  \caption{\textbf{(a)} Baseline bend (BL, $\gamma=0.2$) \textbf{(b)} optimised bend (OPT, variable $\gamma$ with $\gamma_{\max}\!\approx\!0.8$). Frenet--Serret frame with $\hat{\mathbf{s}}$: tangent (streamwise); $\hat{\mathbf{r}}$: radial (centrifugal); $\hat{\mathbf{y}}$: binormal (lateral) unit vectors. Streamwise position is given by the non-dimensional arc length $S=s/D$, where $s$ is the dimensional arc length along the pipe centreline and $D$ is the baseline pipe diameter.}
  \label{fig:coordSys}
\end{figure}

The present work is motivated by two open issues. Firstly, the absence of evidence for relaminarisation in curved pipes at high Reynolds numbers, and secondly, the largely unexplored influence of cross-sectional shape on the Dean vortices and turbulence. We address both of these aspects by investigating whether relaminarisation of turbulent flow can be achieved through geometric modification of the pipe, thereby unlocking the potential for lower drag forces in general piping systems. We consider the combined modification of the local centreline curvature and the cross-sectional geometry. Rather than performing a parametric study of local curvature and cross-sectional profile within a high-dimensional geometric space, we pose and solve an automatic optimisation problem to determine the required changes in these two geometric variables. The optimisation was carried out for the baseline $180^\circ$ bend (BL) at $\mathrm{Re}_D=10,000$ with $\gamma=0.2$, as illustrated in figure~\ref{fig:coordSys}(a). This choice was motivated by earlier work reporting partial inner-wall relaminarisation for \enquote{strong} curvature $\gamma=0.1$ \citep{Noorani2013_IJHFF}. Therefore, we increased the curvature to $\gamma=0.2$ to investigate whether a stronger curvature would further promote relaminarisation in the baseline geometry. We use a free-form shape-optimisation framework based on the steady incompressible Reynolds-Averaged Navier–Stokes (RANS) equations solved with the incompressible solver in \textsc{SU2} \citep{Economon2016} and the $k$--$\omega$ SST closure model. Gradient information is obtained with a discrete-adjoint approach using algorithmic differentiation \citep{Albring2016_AIAA,Sagebaum2019_TOMS}. The objective function to be minimised is the total entropy generation in the flow, which serves as a proxy for the appearance of turbulence. For incompressible and isothermal flows, entropy generation is due to viscous dissipation, and our objective function can be defined as
\begin{equation}
    J = \int_{\Omega} \Phi \mathrm dV
    =  \int_\Omega \left( \tau : \nabla U + \rho \varepsilon \right)  \mathrm d V ,
    \label{eqn:objectiveFunction}
\end{equation} 
where $\tau $ is the mean viscous stress tensor, $\nabla U$ the mean velocity gradient, and $\varepsilon$ denotes the modelled turbulent dissipation obtained from the $k$--$\omega$ SST model. Here, ``$:$'' denotes the double inner product. 
With this objective function, we minimise irreversible losses in the flow \citep{Kock2004}, thus biasing the search toward low-dissipation laminar states. The optimisation region was restricted to the segment between arc lengths $S=3$ and $S=21$. The remaining geometry ($S<3$ and $S>21$) was kept fixed. This is the only imposed optimisation constraint, and is meant to keep the optimised and baseline configurations directly comparable while enabling a practical pressure-loss experiment with measurement stations located in the unchanged straight sections upstream and downstream. After 50 automatic design iterations, the gradient of the objective function vanishes, and a local minimum is reached. The resulting optimal design (OPT) is shown in figure~\ref{fig:coordSys}(b).

As the optimisation framework is based on RANS equations, we perform DNS to verify whether these geometric modifications can achieve relaminarisation. We then analyse the resulting optimal flow to identify the underlying physical mechanisms that relaminarise the flow using DNS results. The DNS was performed with the spectral-element solver $\mathrm{Nek5000}$ \citep{Fischer2008_Nek5000} using the $\mathrm{P}_N$--$\mathrm{P}_{N-2}$ formulation with $\mathrm{7^{th}}$-order polynomials. Time integration was performed with a third-order semi-implicit backward differentiation scheme ($\mathrm{BDF3}$). For both the baseline (BL) and optimised (OPT) geometries, simulations were performed at $\mathrm{Re}_D=10,000$ and $20,000$ on meshes of comparable resolution to \cite{Hufnagel2018} comprising approximately $6.3\times10^{5}$ elements at $\mathrm{Re}_D=10,000$ and $4.7\times10^{6}$ elements at $\mathrm{Re}_D=20,000$. All simulations employ fully-developed turbulent inflow conditions as described in \cite{Hufnagel2018}.
\begin{figure}[t]
  \centering
  \begin{overpic}[width=0.39\linewidth]{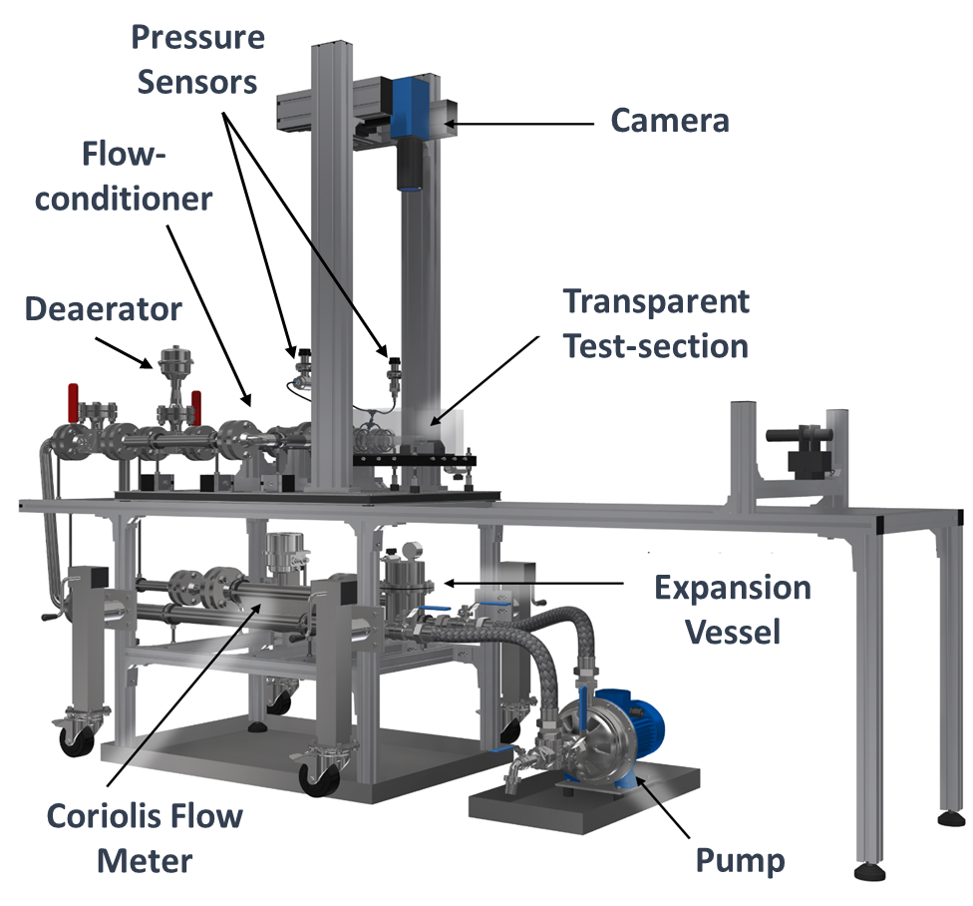}
    \put(35,2){\small\bfseries (a)}
  \end{overpic}\hfill
  \begin{overpic}[width=0.55\linewidth]{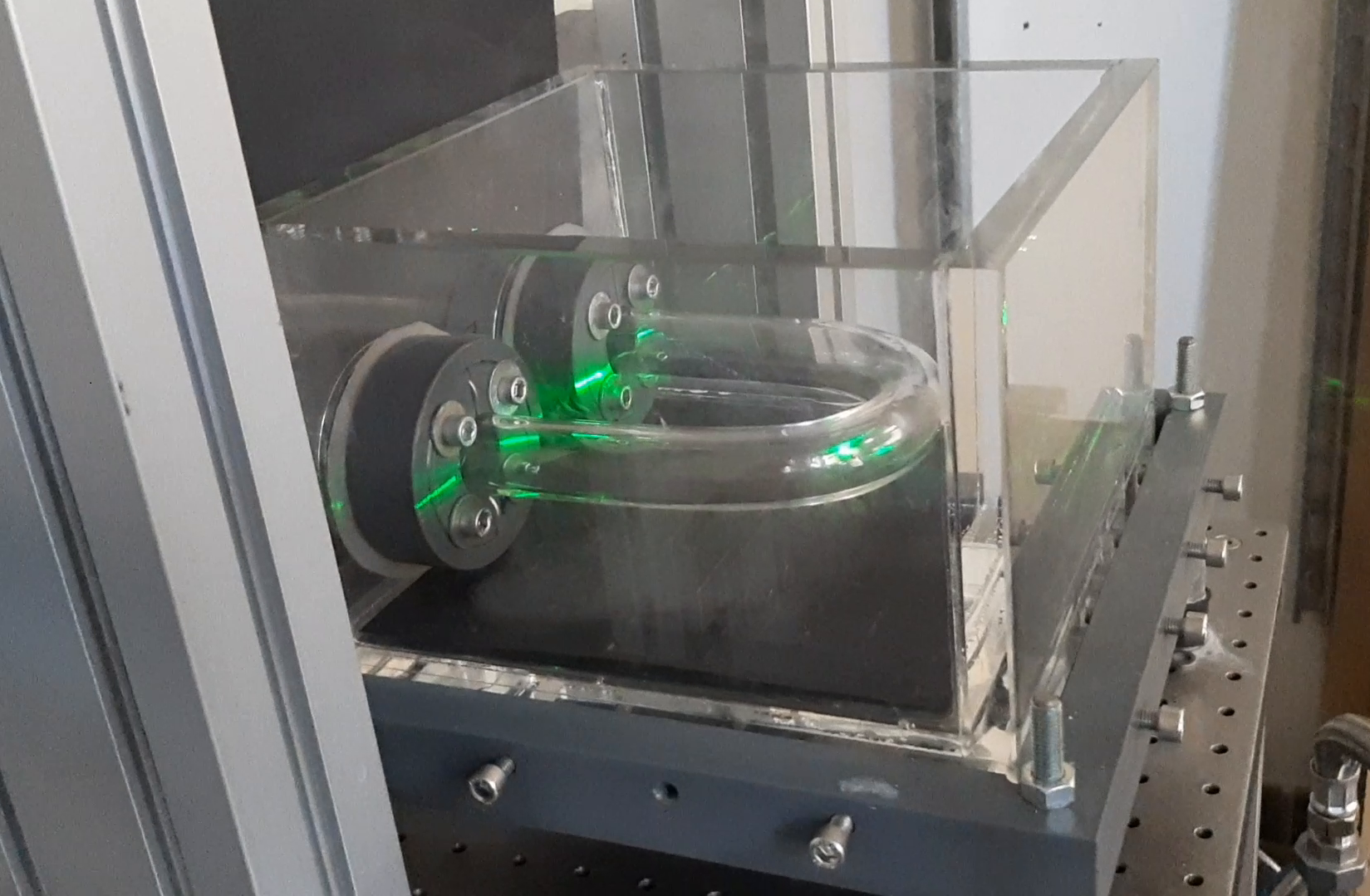}
    \put(-8,2){\small\bfseries (b)}
  \end{overpic}
  
  \caption{\textbf{(a)} Various components of the test-rig and \textbf{(b)} the transparent test section.}
  \vspace{-4mm} 
  \label{fig:testRig}
\end{figure}

Furthermore, the DNS results were validated against pressure loss measurements. The experimental setup is shown in figure~\ref{fig:testRig}, in which water is 
continuously circulated using a radial pump. The pump-induced pulsations are attenuated using a membrane-type expansion vessel connected via flexible hoses, and the flow rate (Reynolds number) is monitored with a Coriolis flowmeter upstream of the test section. Downstream of the flowmeter, a deaeration unit is used to remove entrained gas before entering a modular flow-conditioning section. Pressure measurements are obtained from two differential pressure transducers connected to circumferential wall taps that are combined into ring manifolds at the inlet and outlet of the test section. These measurement stations correspond to $S=0$ and $S=24$ in figure~\ref{fig:coordSys}. Both optimised and baseline geometries were manufactured using vacuum casting with two-component water-clear polyurethane resin. The resin is mixed and degassed under vacuum and poured into a preheated silicone mould, and is then oven-cured, producing an optically transparent geometry. More detailed information about the test-rig can be found in \citet{Baumeister2018_MA}.
\vspace{-5mm} 
\section{Results and discussion}

\begin{figure}[h]
  \centering
  \begin{overpic}[width=0.99\textwidth]{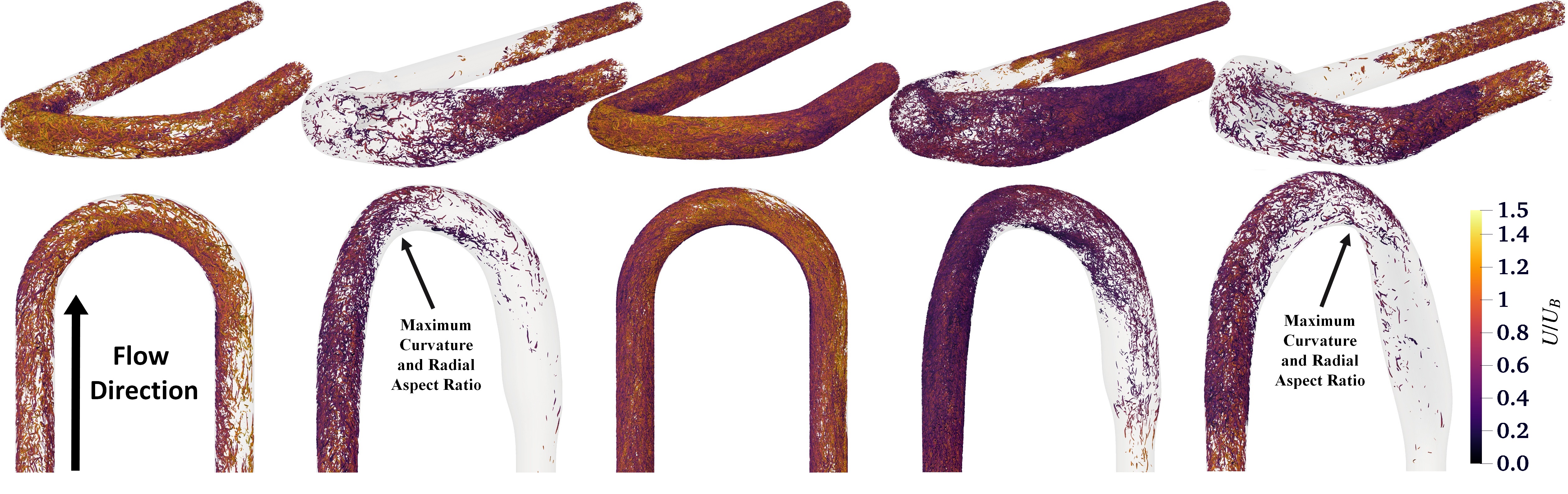}
    \put(3.0,-2){\sffamily\footnotesize\bfseries \textbf{(a) BL 10k}}
    \put(21,-2){\sffamily\footnotesize\bfseries \textbf{(b) OPT 10k}}
    \put(41.5,-2){\sffamily\footnotesize\bfseries \textbf{(c) BL 20k}}
    \put(59.6,-2){\sffamily\footnotesize\bfseries \textbf{(d) OPT 20k}}
    \put(78.5,-2){\sffamily\footnotesize\bfseries \textbf{(e) ALT 10k}}
  \end{overpic}%
  \vspace{1mm} 
  \hfill
  \caption{Isosurfaces of $\lambda_2 = -10~{U_B^2}/{D^2}$, coloured by velocity magnitude.
  \textbf{(a,c)} show the baseline design at $\mathrm{Re_D}=10,000$ and $\mathrm{Re_D}=20,000$,
  respectively. \textbf{(b,d)} In the optimised case at both Reynolds numbers, turbulence
  decays shortly after the bend entry at $S\approx 10$ up to the location where the cross
  section is constrained back to a circular shape at $S\approx 21$. \textbf{(e)} shows another optimised shape from a different local minimum.}
  \vspace{-3mm}
  \label{fig:lambda2}
\end{figure}

The DNS results confirm that the modified geometry attains near full relaminarisation at both Reynolds numbers. Figure~\ref{fig:lambda2}(b) and (d) qualitatively illustrate that turbulent structures convected from the inflow region begin to decay near $S \approx 10$ in the modified design (see the baseline design in figure~\ref{fig:lambda2}(a) and (c) for comparison). Note that both Reynolds numbers are well above the linear stability limit \citep{Massaro2023_PRFluids_180bend}. Moreover, the optimisation results in multiple families of solutions, each converging to a distinct local minimum. For instance, figure~\ref{fig:lambda2}(e) shows an alternative modified geometry at $\mathrm{Re_D}=10,000$ (ALT 10k). This indicates that the geometric modification required to achieve relaminarisation is not unique.  Despite clear differences, both modified geometries have a key feature in common that relaminarises the flow. In both cases, the maximum local curvature coincides with a large cross-sectional aspect ratio where turbulence starts to decay. For brevity, we do not further discuss geometry ALT, as it breaks equatorial symmetry and would require extending the analysis beyond the no-torsion framework.

\begin{figure}[h]
  \centering
  \begin{overpic}[width=0.48\linewidth]
    {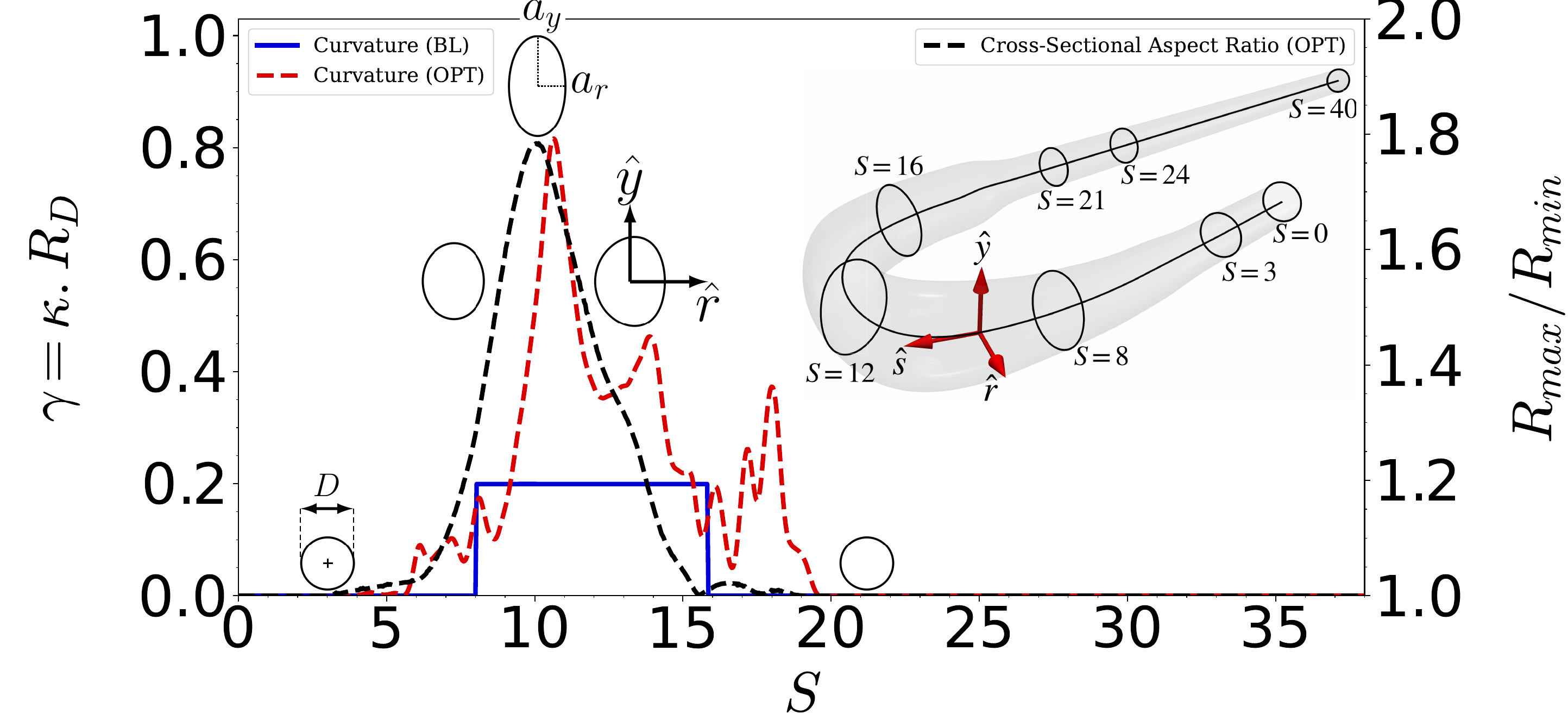}
    \put(0,2){\small\bfseries (a)}
    \end{overpic}
  \begin{overpic}[width=0.5\linewidth]
    {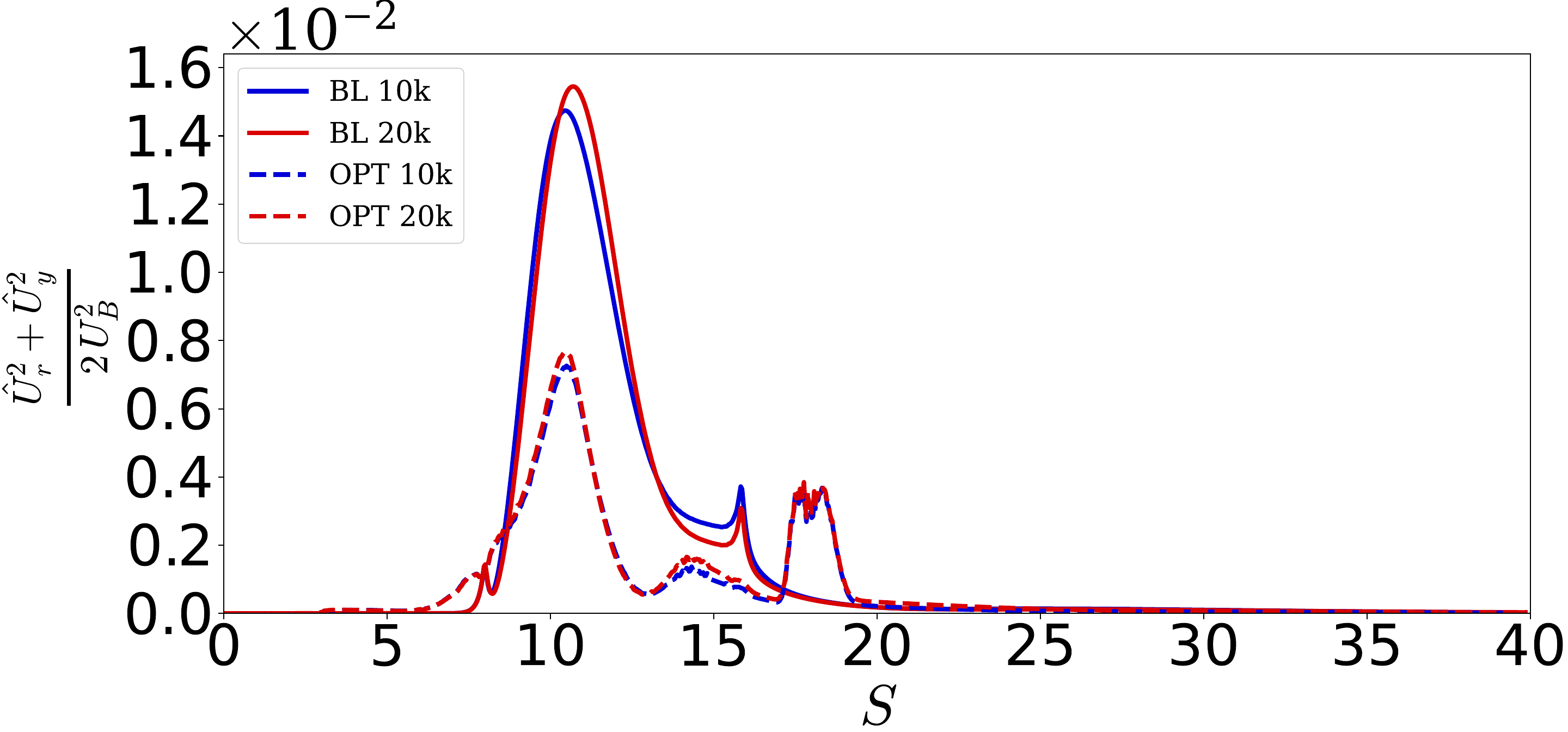}
    \put(2,2){\small\bfseries (b)}
  \end{overpic}
  \caption{(\textbf{a}) Geometric characteristics of baseline and optimised bends: 
  The left axis shows non-dimensional curvature, and the right axis the cross-sectional aspect ratio.
  (\textbf{b}) Kinetic energy of the secondary flow. Despite a stronger curvature, the modified shape has a weaker secondary flow owing to its cross-sectional shape.}
  \label{fig:geoParam_ke}
\end{figure}

Figure~\ref{fig:geoParam_ke}(a) shows the centreline curvature and the cross-sectional aspect ratio of the modified and baseline geometries and highlights two major characteristics of the modified shape. Firstly, the dimensionless curvature parameter $\gamma$ significantly increases from 0.2 in BL to 0.82 in OPT at $S\approx 10$, consistent with previous findings \citep{Huettl2000a,Noorani2013_IJHFF,Rinaldi2019} on the stabilising effect of strong streamwise curvature. Secondly, the cross-sectional profile has an oval shape elongated in the binormal direction $\hat{y}$ with the ratio of its principal axes reaching a maximum of $R_{\text{max}}/R_{\text{min}} = 1.78$ near $S \approx 10$. Note that both the curvature and the cross-sectional aspect ratio reach their maximum nearly simultaneously at $S \approx 10$. We analyse the impact of these two geometric modifications by deriving the Reynolds-averaged equations in the Frenet–Serret coordinate system shown in figure~\ref{fig:coordSys} (see Appendix~\ref{app:FS} for the equations). We adopt the Frenet--Serret reference frame with unit vectors $\hat{\mathbf{s}}$ (tangent/streamwise), $\hat{\mathbf{r}}$ (radial/centrifugal), and $\hat{\mathbf{y}}$ (binormal/lateral), rather than the common toroidal coordinate system because the mean flow lacks azimuthal homogeneity, and flow dynamics in the radial and binormal directions differ fundamentally.
We report the streamwise locations using the non-dimensional arc length $S=s/D$, where $s$ is the dimensional arc length and $D$ is the pipe diameter.
\begin{figure}[h]
  \centering
  
  \begin{overpic}[width=0.485\linewidth]
    {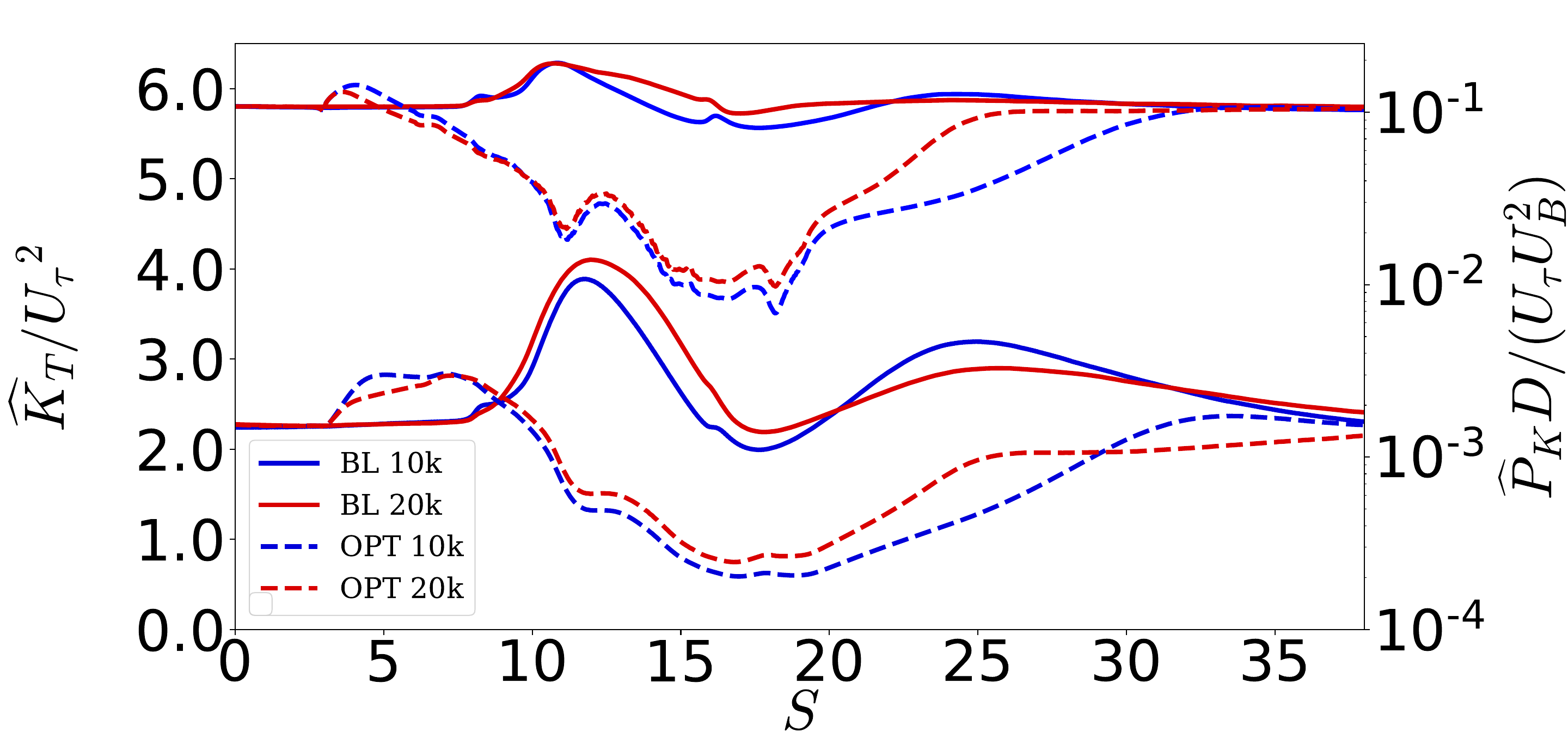}
    \put(0,2){\small\bfseries (a)}
    \put(79.5,29.5){\color{black}\scriptsize\bfseries $\widehat P_K$}
    \put(81.5,34){\color{black}\scriptsize\vector(0,1){5}}
    \put(17.5,26){\color{black}\scriptsize\bfseries $\widehat K_T$}
    \put(19,25){\color{black}\scriptsize\vector(0,-1){5}}
  \end{overpic}

  \begin{overpic}[width=0.495\linewidth]
    {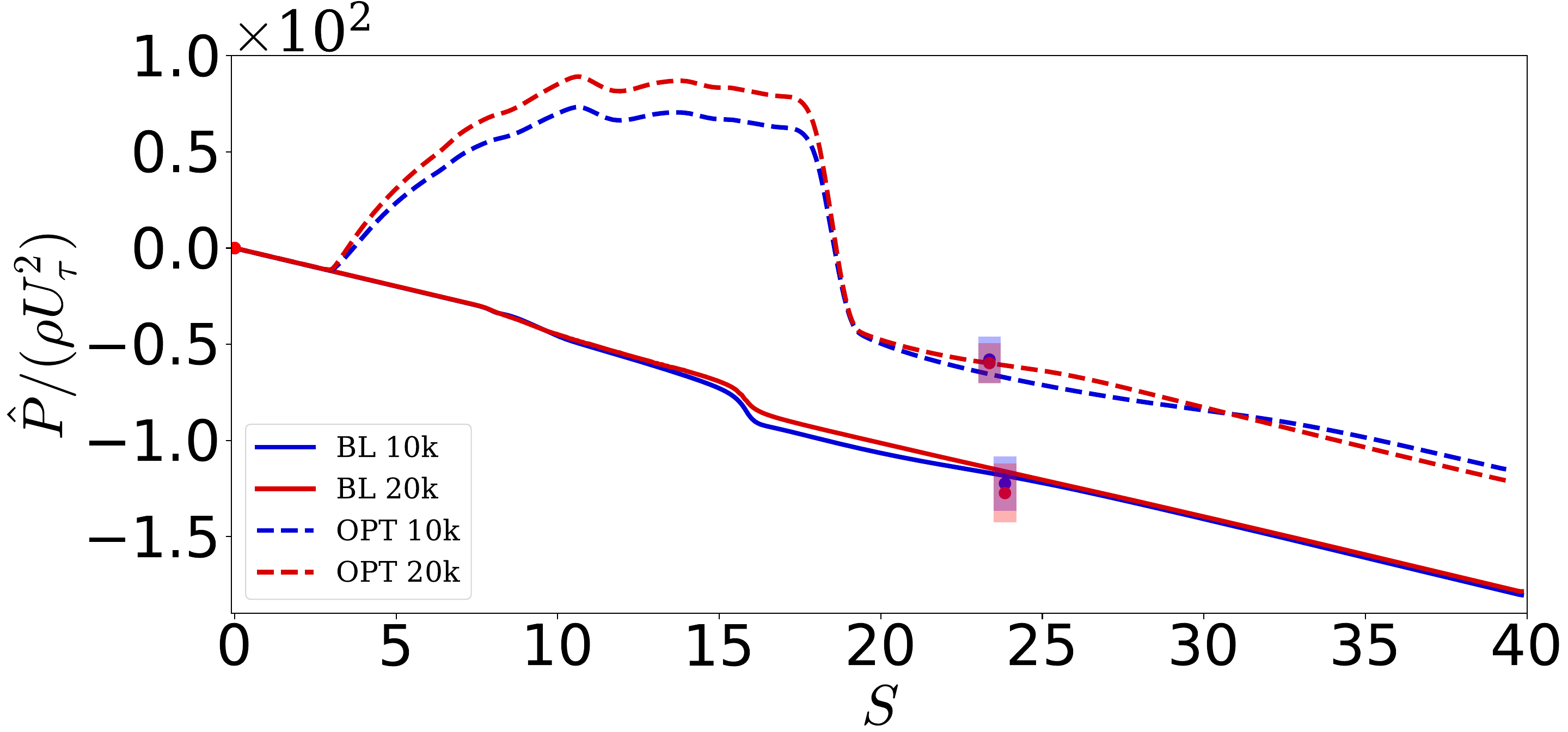}
    \put(60,35){\color{black}{\fontsize{5}{5}\selectfont\bfseries Experiment}}
    \put(70,34){\color{black}\scriptsize\vector(-2,-3){5.5}}
    \put(70,34){\color{black}\scriptsize\vector(-1,-3){5.3}}
    \put(2,2){\small\bfseries (b)}
  \end{overpic}
  \caption{\textbf{(a)} Cross-sectionally averaged turbulent kinetic energy $\widehat{K}_T$ and its production
  $\widehat{P}_K$. \textbf{(b)} Streamwise mean pressure for BL and OPT at $\mathrm{Re_D}=10,000$ and $20,000$. The friction velocity $U_\tau$ corresponds to that of fully developed turbulent flow at the inflow $S=0$. Markers indicate pressure loss measurements, and shaded bands denote one standard deviation ($\pm\sigma$), accounting for turbulent fluctuations and measurement uncertainty.}
  \label{fig:tke_pressure}
  \vspace{-2mm} 
\end{figure}
In this frame, we define the cross-sectional average of a time-averaged field 
$g(r,y,s)$ as

\begin{equation}
\widehat{g}(s):=\frac{1}{|\mathcal{A}(s)|}\iint_{\mathcal{A}(s)} g(r,y,s)\,\mathrm{d}r\,\mathrm{d}y,
\qquad
|\mathcal{A}(s)| := \iint_{\mathcal{A}(s)} \mathrm{d}r\,\mathrm{d}y
\end{equation}
where $\mathcal A(s)$ denotes the pipe's cross section in the $(r,y)$ plane at arc length $s$. Expressing all terms in this frame enables us to isolate and study the geometric effects. Let us consider $P_{ss}$, the production term of the streamwise Reynolds stress $R_{ss}=\overline{u'_s u'_s}$ in the Frenet--Serret frame. In this curvilinear system, curvature adds extra metric contributions, which we group into $P_{ss}$ as follows.
\vspace{-1mm} 
\begin{equation}
P_{ss} = -2\Bigg[
R_{ss}\left(\frac{\partial_s U_s + \kappa\,U_r}{h_s}\right)
+ R_{sy}\left(\partial_y U_s\right)
+ R_{sr}\left(\partial_r U_s + \frac{\kappa\,U_s}{h_s}\right)
\Bigg],
\label{eq:Pss}
\end{equation}
Here, $U_i$ and $R_{ij}=\overline{u'_i u'_j}$ denote the mean velocity and Reynolds-stress components, respectively. The dimensional centreline curvature $\kappa=1/R_c$ (related to the non-dimensional curvature parameter $\gamma$ by $\kappa=2\gamma/D$) appears explicitly in the production of the streamwise Reynolds stress. The strength of these curvature terms is determined by the local curvature $\kappa/h_s$ (with $h_s=1+\kappa r$), being strongest near the inner wall and weakest near the outer wall. As fluid elements accelerate toward the outer wall under the centrifugal force, the radial velocity satisfies $U_r \geq0$ everywhere inside the bend except near the top and bottom walls. Consequently, the first curvature term
\begin{equation}
\frac{\kappa}{h_s} R_{ss} U_r ,
\label{eq:Pss_kappa_term}
\end{equation}
in Eq.~\eqref{eq:Pss} has a stabilising effect on the streamwise Reynolds stress $R_{ss}$ by reducing its production everywhere, except in the thin regions in the vicinity of the upper and lower walls where the secondary flow returns towards the inner wall. Moreover, the shear-production term $R_{sr}\partial_r U_s$ is modified by the additional curvature contribution in Eq.~\eqref{eq:Pss}:
\vspace{-1mm} 
\begin{equation}
R_{sr}\Big(\partial_r U_s + \frac{\kappa\,U_s}{h_s}\Big).
\label{eq:Pss_kappa_term_shear}
\end{equation}
This curvature term also appears in the production of radial stresses, with opposite sign and twice the magnitude (see Eq.~\eqref{eq:AppendPrr}). It locally exchanges energy between the streamwise and radial Reynolds stresses through the radial-streamwise shear stress and makes a net contribution to the production of turbulent kinetic energy (TKE). On the inner wall, it enhances the streamwise stress while reducing the radial stress, and its net effect suppresses the TKE production (see Eq.~\eqref{eq:AppendPk}). This curvature term has the opposite effect on the outer wall and is responsible for the localisation of the TKE production toward the outer wall (see \cite{Rinaldi2019}) and the inner wall partial relaminarisation observed in \cite{Noorani2013_IJHFF}. At the same time, the curvature induces the unstable Dean vortices, contributing to the production of cross-stream Reynolds stresses. 
Moreover, the secondary flow redistributes the TKE produced on the outer wall by transporting turbulent structures into the core and toward the inner wall. Therefore, the streamwise flow curvature has both suppressing and enhancing effects on turbulence, and the balance between these competing mechanisms determines the net outcome.  

\begin{figure}[h]
  \centering
  \begin{overpic}[width=.95\linewidth,trim={0 0 0 1},clip]{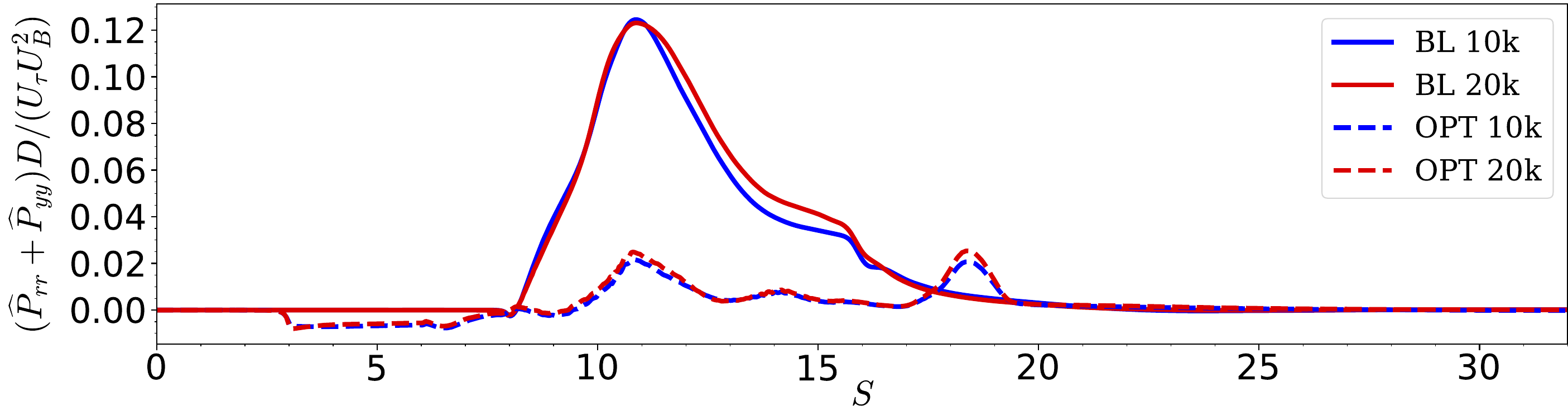}
  \end{overpic}
   \vspace{-3mm} 
  \caption{Production of cross-stream Reynolds stresses. In contrast to the optimised design, the unstable Dean vortices in the baseline design sustain the turbulence inside the bend. 
  }
    \vspace{-1.5mm} 
  \label{fig:P_crossStream}
\end{figure}

Figure~\ref{fig:tke_pressure}(a) shows that the TKE production initially increases at $S=4$, intensifying the TKE from $S=4$ to $S=8$ in the optimised design. This is in line with the observation by \cite{Kuehnen2018} that the return to laminar motion is accomplished by initially increasing turbulence intensities. Following that, the production decreases exponentially from $S=5$ to $S=18$ at $\mathrm{Re_D}=10,000$ (OPT 10k) and $\mathrm{Re_D}=20,000$ (OPT 20k). As a consequence, the TKE steadily decays from $S=8$ to $S=18$. In contrast, the TKE production inside the baseline bend increases, resulting in $1.7$ to $1.8$ times higher peak TKE compared to the straight section. This is mainly due to the instability of the secondary flow, which generates cross-stream Reynolds stresses ($R_{yy}$ and $R_{rr}$) and raises the overall turbulence intensity.
Here, $R_{yy}=\overline{u'_y u'_y}$ and $R_{rr}=\overline{u'_r u'_r}$ denote the binormal and radial normal Reynolds stresses in the Frenet--Serret frame. In straight pipes, these two components arise primarily through redistribution by the pressure–strain, whereas in curved pipes, they are directly amplified by the production terms $P_{yy}$ and $P_{rr}$ (see Eq.~\eqref{eq:AppendPyy}, \eqref{eq:AppendPrr} and figure ~\ref{fig:P_crossStream}).

The increase in cross-stream Reynolds stresses intensifies the shear-stress production $P_{sy}$ and $P_{sr}$ through $R_{yy} \partial_y U_s$ and $R_{rr} \partial_r U_s$ (see Eq.~\eqref{eq:shearProd_sy} and \eqref{eq:shearProd_sr}). The intensified shear stresses feed back into the production of streamwise stress $R_{ss}$ through $R_{sy}\partial_y U_s$ and $R_{sr}\partial_r U_s$
(see Eq.~\eqref{eq:Pss}), countering the stabilising curvature effects, thereby regenerating streaks and sustaining turbulence. This mechanism explains why previous studies have not reported relaminarisation at higher Reynolds numbers merely through curvature modifications without changing the cross section. In other words, a curvature increase alone tends to strengthen the Dean vortices and the associated cross-stream Reynolds stresses, and relaminarisation requires a concurrent cross-sectional modification to counter this effect. Our complementary wall-resolved large-eddy simulations of circular bends, matched to the optimal local curvature and local Dean number, confirm the absence of relaminarisation when the cross section remains circular. The mean streamwise vorticity equation is given in Eq.~\eqref{eq:Ws}. We neglect streamwise variations of the mean flow and curvature ($\partial_s(\cdot)\approx 0$ and $\kappa'(s) \approx 0$) and keep the leading-order term as an interpretive
scaling guide. 
\vspace{-1mm} 
\begin{equation}
\frac{D\omega_s}{Dt}
= 
{\partial_y\!\left(\frac{\kappa}{h_s}\,U_s^2\right)}
\;+\;
{\frac{\kappa}{h_s}\,U_r\,\omega_s}
\;+\;
\nu\,(\nabla^2\omega)_s
\;-\;
\bigl[\nabla\times(\nabla\cdot \mathbf{R})\bigr]_s
\label{eq:Ws}
\end{equation}
As shown in Eq.~\eqref{eq:Ws}, the binormal gradient of the centrifugal force 
\vspace{-1mm} 
\begin{equation}
\partial_y\!\left(\frac{\kappa}{h_s} U_s^2\right)\sim \frac{1}{a_y} \frac{\kappa \hat{U_s}^2}{h_s},
\label{eq:vorticity_rhs}
\end{equation}
is the main driving term of the Dean vortices. Here, $a_y$ is the local lateral half-width of the pipe cross section in the binormal direction $\hat{\mathbf{y}}$ (see figure~\ref{fig:geoParam_ke}(a)), and $\mathbf{R}$ denotes the Reynolds-stress tensor. Hence, increasing the curvature without modifying the cross-sectional profile results in stronger Dean vortices. The instability of the Dean vortices (see supplementary movie 1), in turn, sustains the turbulence inside the bend through cross-stream Reynolds stresses. Near the inner wall, the major principal axes of both the Reynolds stress and production tensors become closely aligned with the binormal direction, and $R_{yy}$ emerges as the primary carrier of turbulence. 

\begin{figure}[h]
  \centering
  \begin{overpic}[width=.95\linewidth,trim={1 130 11 148},clip]{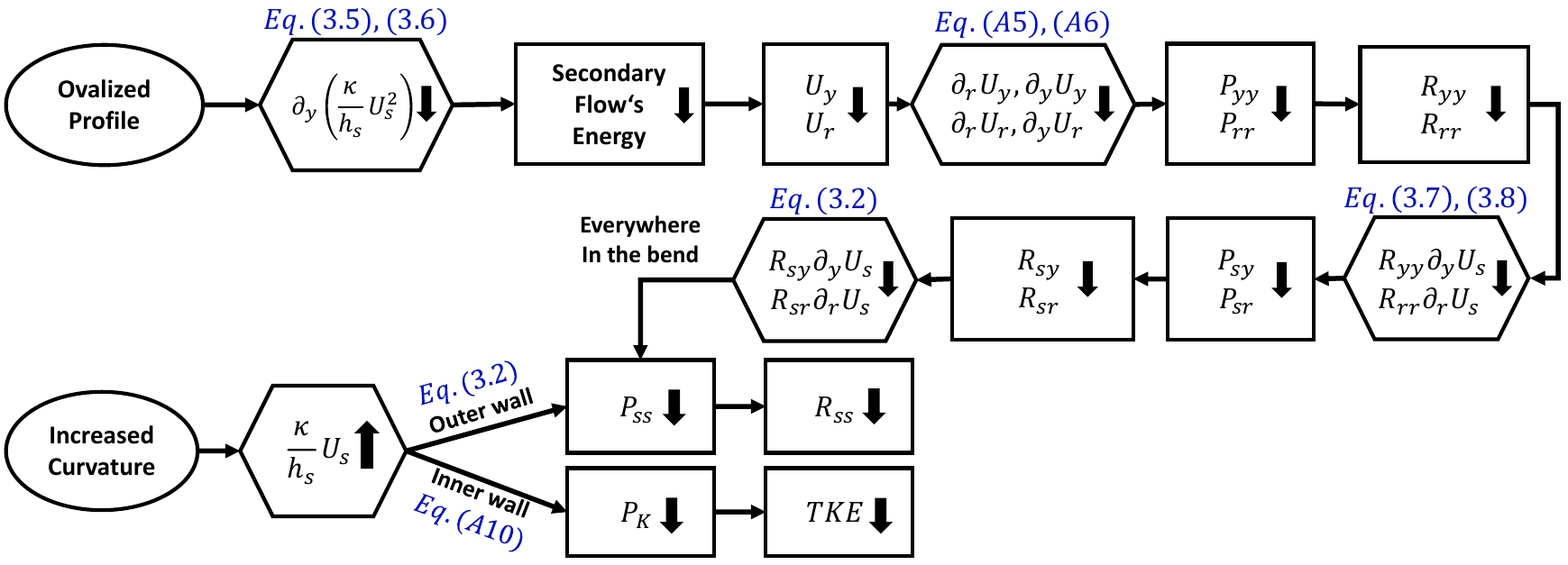}
  \end{overpic}
   \vspace{-3mm} 
   \caption{Schematic illustration of how geometric features (streamwise curvature and an oval cross-sectional profile) affect turbulence, demonstrating the relaminarisation mechanism in the optimised bent pipe.}
    \vspace{-1.5mm} 
  \label{fig:mechanism_chart}
\end{figure}

The optimised design counters this effect by ovalising the cross section and increasing the lateral half-width $a_y$ as curvature increases (see figure~\ref{fig:geoParam_ke}(a)). Enlarging the cross-sectional area reduces the averaged streamwise velocity $\hat{U_s}$, and together with the increase in the lateral half-width  $a_y$ weakens the driving term of the unstable Dean vortices in Eq.~\eqref{eq:vorticity_rhs}. As a direct consequence, figure~\ref{fig:geoParam_ke}(b) shows a 50\% reduction in the peak kinetic energy of the secondary flow relative to the baseline geometry. By reducing the energy of the Dean vortices, cross-stream Reynolds stress production terms diminish as shown in figure~\ref{fig:P_crossStream}, and consequently $R_{yy}$ and $R_{rr}$ drop. The leading terms in shear-stress productions are 
\begin{align}
P_{sy} &\approx -R_{yy}\partial_y U_s - R_{yr}\partial_r U_s, \label{eq:shearProd_sy}\\
P_{sr} &\approx -R_{yr}\partial_y U_s - R_{rr}\partial_r U_s + \frac{\kappa U_s}{h_s}\left(2R_{ss}-R_{rr}\right). \label{eq:shearProd_sr}
\end{align}
With scaling $\partial_y U_s \sim \hat{U_{s}}/a_y$, the increase in $a_y$ due to ovalisation reduces the binormal mean velocity gradient in the production of shear stresses without affecting the curvature factor $\kappa/h_s$. 
Reducing cross-stream stresses ($R_{yy}$ and $R_{rr}$) and the binormal mean velocity gradient $\partial_y U_s$ suppresses the shear-stress productions in Eq.~\eqref{eq:shearProd_sy} and ~\eqref{eq:shearProd_sr}. As shear stresses $R_{sr}$ and $R_{sy}$ are subdued, the streamwise production $P_{ss}$ in Eq.~\eqref{eq:Pss} drops, and the near-wall streaks diminish (see figure~\ref{fig:mechanism_chart} for a schematic summary of this mechanism).

The combined effect of increased streamwise curvature and the weaker secondary motion suppresses turbulence  (see supplementary movie 2) and reduces the mean velocity gradient. Consequently, the optimised bend achieves a 53\% reduction in pressure loss relative to the baseline bend, confirmed by experiment in figure~\ref{fig:tke_pressure}(b). Due to relaminarisation, its pressure loss is 36\% below that of a fully developed straight turbulent pipe of equal length at the same Reynolds number. Furthermore, based on our pressure loss measurements, this behaviour persists beyond $\mathrm{Re_D}=20,000$. However, pressure loss data alone is not conclusive evidence of relaminarisation. As shown in figure~\ref{fig:tke_pressure}(a), the primary difference in the TKE and TKE production of OPT 10k and OPT 20k is in the re-transition region beyond $S\approx 20$, where the geometry returns to a straight circular pipe due to the optimisation constraint. At $Re_D=20,000$, the residual turbulent slugs near $S\approx 20$ grow more rapidly, and turbulence recovers faster in the straight section of the pipe (see figure~\ref{fig:lambda2}(d) and figure~\ref{fig:tke_pressure}(a)). Nevertheless, in the relaminarising region from $S= 8$ to $S= 18$, the TKE and TKE production remain largely unchanged between OPT~10k and OPT~20k, indicating only a weak Reynolds-number dependence (see figure~\ref{fig:tke_pressure}(a)). It is worth noting that, in addition to the primary mechanisms discussed above, the optimised geometry also benefits from two secondary effects that further reduce the objective function. First, the favourable pressure gradient near the bend exit contributes to the stabilisation of the remaining turbulent patches on the outer wall. Second, the cross-sectional area increases in the modified bend, and the flow momentarily reaches a minimum local Reynolds number of approximately $Re_{D}^{\mathrm{loc}}\approx 0.65\,Re_D$ corresponding to $Re_{D}^{\mathrm{loc}}\approx 13,000$ for OPT~20k (see Appendix~\ref{app:extraData}). This increase in area reduces the mean velocity gradient and contributes to the reduction in the frictional losses, but it is not sufficient to induce relaminarisation at sufficiently high Reynolds numbers. We simulated the flow in a straight pipe with a local cross-sectional area equal to that of our optimised geometry, and observed a $32\%$ reduction in pressure loss due to the local area increase. However, relaminarisation effects are absent due to the lack of curvature-induced mechanisms. In the curved case, an increase in cross-sectional area alone may reduce frictional losses, but relaminarisation requires suppression of the Dean vortices, which necessitates reducing the centrifugal forcing term in Eq.~\eqref{eq:Ws} by adopting an oval cross section.

\vspace{-2mm} 
\FloatBarrier
\vspace{-2mm} 
\section{Overview and conclusions}

The present work shares with \cite{Kuehnen2018} the main idea that turbulence can be suppressed by interrupting the self-sustaining near-wall cycle, but achieves it passively by geometric modifications. The cycle entails the following: streamwise rolls lift up the streaks, the streaks become unstable and break down, and the resulting fluctuations regenerate the rolls \citep{Hamilton1995,Jimenez1999,Waleffe1997}. We disrupt this loop in the following way. Firstly, the strong streamwise curvature has a suppressing influence on the TKE production on the inner wall and reduces the production of streamwise Reynolds stress on the outer wall, hence the streaks are no longer reinforced. Secondly, an oval cross section elongated in the binormal direction has two additional effects. By increasing the cross-sectional area and the lateral half-width, the centrifugal force and the centrifugal production of the secondary flow in the streamwise mean vorticity equation are reduced. This, in turn, weakens the unstable Dean vortices and decreases their energy feedback into turbulence via the cross-stream Reynolds stresses. In addition, through attenuation of the binormal mean shears and the cross-stream Reynolds stresses, it reduces the shear-stress productions and consequently weakens the shear stresses that reinforce the streaks. 
As both the direct energy and the feedback mechanisms are disrupted, streaks are suppressed, and turbulence collapses inside the bend. 

The present work opens new avenues to control turbulence in curved pipe flow and thus reduce frictional losses at the walls. Given the novel insights into the governing equations and the turbulent kinetic-energy budget, and their connection to both pipe curvature and cross-sectional shape, new approaches in other curved wall-bounded flows are also possible.

\begin{bmhead}[Funding.]
This project was partially funded by the Bavarian Research Foundation (project AZ-1232-16).
This work has also received funding from the European High Performance Computing Joint Undertaking (JU), together with Sweden, Germany, Spain, Greece and Denmark, under grant agreement no.~101093393.
The computing hardware was funded by the German Research Foundation (DFG).
\end{bmhead}

\begin{bmhead}[Acknowledgements.]
The authors gratefully acknowledge the scientific support and high-performance computing resources
provided by the Erlangen National High Performance Computing Center (NHR@FAU) of the
Friedrich--Alexander--Universit\"at Erlangen--N\"urnberg (FAU).
\end{bmhead}

\begin{bmhead}[Declaration of interests.]
The authors report no conflict of interest.
\end{bmhead}

\vspace{-3mm} 
\FloatBarrier
\vspace{-3mm} 
\appendix
\section{Reynolds-stress production and mean vorticity transport equations}
\phantomsection
\label{app:FS}

\setcounter{equation}{0}
\renewcommand{\theequation}{A\arabic{equation}}

We use the orthogonal centreline coordinate system $(s,y,r)$ shown in figure~\ref{fig:coordSys} with curvature $\kappa=\kappa(s)$ and no torsion. The orthonormal basis vectors $\{\hat{\mathbf{s}},\hat{\mathbf{y}},\hat{\mathbf{r}}\}$ satisfy
\begin{equation}
\partial_s\hat{\mathbf{s}}=-\kappa\,\hat{\mathbf{r}},\qquad
\partial_s\hat{\mathbf{r}}=+\kappa\,\hat{\mathbf{s}},\qquad
\partial_q\hat{\mathbf{y}}=0\ \ (q\in\{s,y,r\}).
\label{eq:AppendFS_rel}
\end{equation}
The scale factors are
\begin{equation}
h_s(s,r)=1+\kappa(s)\,r,\qquad h_y=h_r=1.
\label{eq:Appendscale_factors}
\end{equation}
With mean velocity components $U_i$, the continuity equation reads
\begin{equation}
\frac{\partial_s U_s+\kappa\,U_r}{h_s}+\partial_y U_y+\partial_r U_r=0.
\label{eq:Appendcont}
\end{equation}
The Reynolds stress is denoted by $R_{ij}=\overline{u'_i u'_j}$. The curvature generates additional terms through covariant advection. We absorb these curvature terms into the mean-shear
production, regroup and simplify by applying \eqref{eq:Appendcont}. The resulting production terms read:

\begin{align}
P_{ss} &=
-2\Bigg[
R_{ss}\left(\frac{\partial_s U_s + \kappa\,U_r}{h_s}\right)
+ R_{sy}\left(\partial_y U_s\right)
+ R_{sr}\left(\partial_r U_s + \frac{\kappa\,U_s}{h_s}\right)
\Bigg].
\label{eq:AppendPss}\\[1mm]
P_{yy} &=
-2\Bigg[
R_{sy}\left(\frac{\partial_s U_y}{h_s}\right)
+ R_{yy}\partial_y U_y
+ R_{yr}\partial_r U_y
\Bigg]
\label{eq:AppendPyy}\\[1mm]
P_{rr} &=
-2\Bigg[
R_{sr}\left(\frac{\partial_s U_r - 2\kappa\,U_s}{h_s}\right)
+ R_{yr}\partial_y U_r
+ R_{rr}\partial_r U_r
\Bigg]
\label{eq:AppendPrr}\\[1mm]
P_{sy} &=
-\Bigg[
R_{ss} \left(\frac{\partial_s U_y}{h_s} \right)
+ R_{yy}\partial_y U_s
- R_{sy}\partial_r U_r
+ R_{sr}\partial_r U_y
+ R_{yr}\Big(\partial_r U_s + \frac{\kappa\,U_s}{h_s}\Big)
\Bigg]
\label{eq:AppendPsy}\\[1mm]
P_{sr} &=
-\Bigg[
R_{ss}\left(\frac{\partial_s U_r - 2\kappa\,U_s}{h_s}\right)
- R_{sr}\partial_y U_y
+ R_{rr}\Big(\partial_r U_s + \frac{\kappa\,U_s}{h_s}\Big)
+ R_{sy}\partial_y U_r
+ R_{yr}\partial_y U_s
\Bigg]
\label{eq:AppendPsr}\\[1mm]
P_{yr} &=
-\Bigg[
R_{sy}\left(\frac{\partial_s U_r - 2\kappa\,U_s}{h_s}\right)
+ R_{yy}\partial_y U_r
+ R_{yr}\partial_r U_r
+ R_{sr}\frac{\partial_s U_y}{h_s}
+ R_{yr}\partial_y U_y
+ R_{rr}\partial_r U_y
\Bigg]
\label{eq:AppendPyr}
\end{align}
The turbulent kinetic energy production $P_k=\tfrac12 P_{ii}$ is
\begin{align}
P_k
=&-
\Bigg[
 R_{ss}\left(\frac{\partial_s U_s + \kappa\,U_r}{h_s}\right)
+ R_{yy}\partial_y U_y
+ R_{rr}\partial_r U_r
+ R_{sy}\left(\partial_y U_s + \frac{\partial_s U_y}{h_s}\right)
\notag\\[-2pt]
&+ R_{sr}\left(\partial_r U_s + \frac{\partial_s U_r}{h_s} - \frac{\kappa\,U_s}{h_s}\right)
+ R_{yr}\left(\partial_r U_y + \partial_y U_r\right)
\Bigg].
\label{eq:AppendPk}
\end{align}
We derive the mean streamwise vorticity transport equation in this frame. For an incompressible flow, the mean vorticity $\boldsymbol{\omega}=\nabla\times\mathbf{U}$ satisfies
\begin{equation}
\frac{D\boldsymbol{\omega}}{Dt}
=
(\boldsymbol{\omega}\cdot\nabla)\mathbf{U}
+\nu\nabla^2\boldsymbol{\omega}
-\nabla\times(\nabla\cdot \mathbf{R})
\label{eq:Appendvortvec}
\end{equation}
For the streamwise component, one obtains
\begin{align}
(\boldsymbol{\omega}\cdot\nabla\mathbf{U})_s
&=
\frac{\omega_s}{h_s}\big(\partial_s U_s+\kappa U_r\big)
+ \omega_r\,\partial_r U_s
+ \partial_y U_s \left(\partial_r U_s-\frac{1}{h_s}\partial_s U_r\right)
+ \frac{1}{2}\,\partial_y\left(\frac{\kappa U_s^2}{h_s}\right).
\label{eq:AppendWs_full}
\end{align}
To isolate the primary effects of curvature and the cross-sectional profile on secondary flow, we neglect the streamwise variations of the mean flow and curvature, assuming $\partial_s(\cdot)\approx 0$ and $\kappa'(s) \approx 0$ and keep the leading-order term as an interpretive scaling guide. 
%. This assumption holds exactly for fully developed toroidal flows. After moving the curvature term from the covariant advection to the right-hand side, the $s$-component of the mean vorticity transport reads:
% To isolate the primary effects of curvature and the cross-sectional profile on secondary flow, we neglect the streamwise variations of the mean flow and curvature, assuming $\partial_s(\cdot)\approx 0$ and $\kappa'(s) \approx 0$. This approximation is used as an interpretive scaling argument to identify the leading-order terms, and it holds exactly for fully developed toroidal flows. After moving the curvature term from the covariant advection to the right-hand side, the $s$-component of the mean vorticity transport reads:
\begin{align}
\frac{D\omega_s}{Dt}
&=
\underbrace{\partial_y\left(\frac{\kappa}{h_s}U_s^2\right)}_{\text{centrifugal production}}
+\underbrace{\frac{\kappa}{h_s}U_r\,\omega_s}_{\text{radial redistribution}}
+\nu\,(\nabla^2\boldsymbol{\omega})_s
-\bigl[\nabla\times(\nabla\cdot \mathbf{R})\bigr]_s 
\label{eq:AppendWs}
\end{align}
\begin{align}
 \frac{D\omega_s}{Dt}=\partial_t\omega_s
 + \frac{U_s}{h_s}\partial_s \omega_s+\
 U_y\partial_y\omega_s
 +U_r\partial_r\omega_s.
\label{eq:AppendScalarAdvection}
\end{align}
The first term on the right-hand side of Eq.~\eqref{eq:AppendWs} produces mean streamwise vorticity via the binormal gradient of centrifugal force. The second term redistributes the streamwise vorticity through radial transport.

%\vspace{-4mm}
\section{Inflow profiles and streamwise variation of area and bulk Reynolds number}
\phantomsection
\label{app:extraData}%

\begin{figure}[h]
  \centering
  
  \begin{overpic}[width=0.49\linewidth]
    {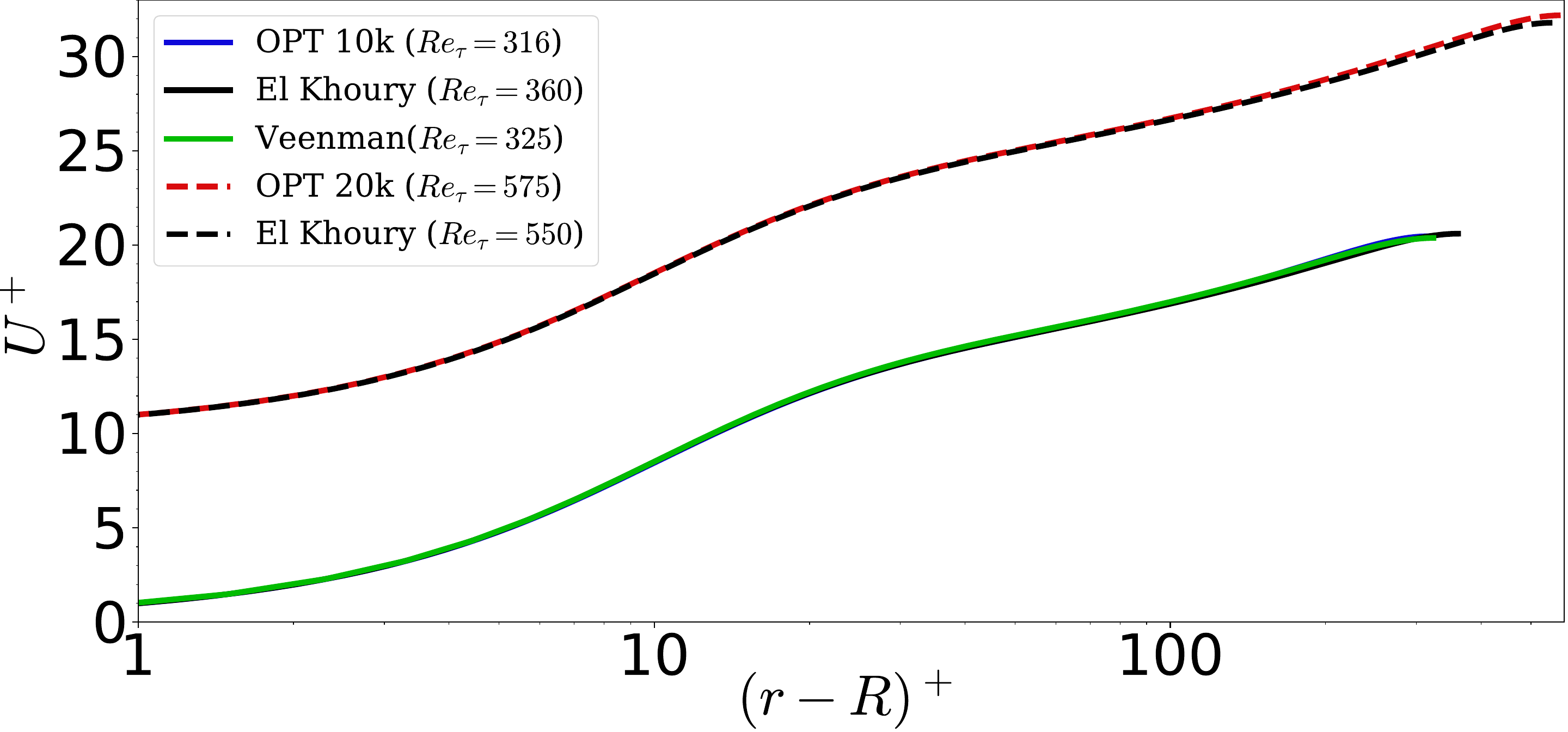}
    \put(6,-2){\small\bfseries (a)}
    \end{overpic}
  \begin{overpic}[width=0.49\linewidth]
    {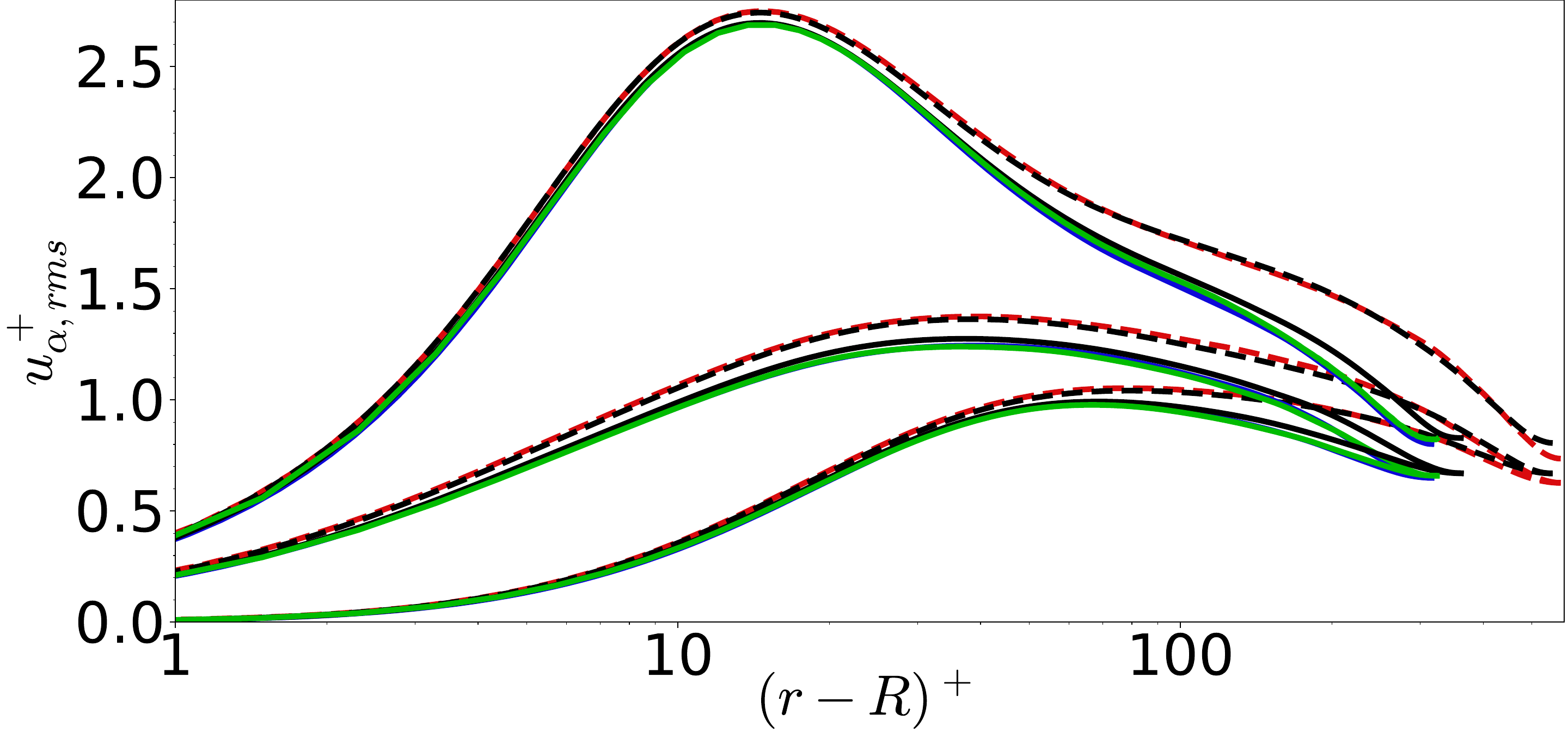}
    \put(8,-2){\small\bfseries (b)}
\put(46,41){\fontsize{7}{8}\selectfont $u_{z,\mathrm{rms}}^+$}
\put(46,28){\fontsize{7}{8}\selectfont $u_{\theta,\mathrm{rms}}^+$}
\put(46,10){\fontsize{7}{8}\selectfont $u_{r,\mathrm{rms}}^+$}
\end{overpic}
\caption{\textbf{(a)} Mean axial velocity profile ($Re_D=20,000$ case shifted by $10 U^+$) and \textbf{(b)} turbulence fluctuations in inner scaling at $S=2$ for the OPT cases in polar coordinates, compared to reference DNS.}
  \label{fig:appendix_profile}
\end{figure}

\begin{figure}[h]
  \centering
  \begin{overpic}[width=.85\linewidth]{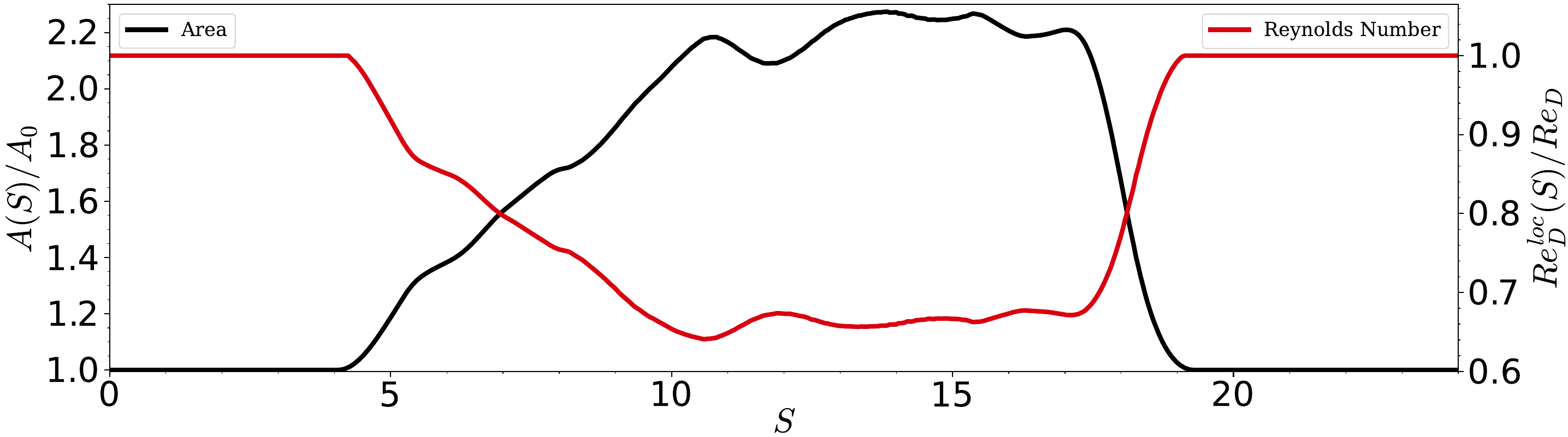}
  \end{overpic}
   \vspace{-1mm} 
   \caption{Local cross-sectional area $A(S)$ non-dimensionalised by the straight-section area $A_0$ (left axis) and local bulk Reynolds number $Re^{\mathrm{loc}}_D(S)$ normalised by the straight-section Reynolds number $Re_D$ (right axis).}
  \label{fig:appendix_Area_Re}
\end{figure}

This appendix provides additional information on the upstream flow statistics and the variation of the cross-sectional area and local Reynolds number in the optimised bend. The statistics were obtained after discarding the initial transient state of the simulation and were averaged over $112$ and $38$ eddy-turnover times $(R/u_\tau)$ for $Re_D=10,000$ and $20,000$ respectively. Since the baseline and optimised geometries are identical for $S<3$, only the optimised cases are shown in figure~\ref{fig:appendix_profile}. The results are compared with reference DNS data from \citet{Veenman2004} and \citet{ElKhoury2013} for fully developed turbulent pipe flow. Figure~\ref{fig:appendix_Area_Re} shows the variation of the cross-sectional area and the corresponding local Reynolds number along the streamwise direction.

% ------------------------------------------------------------------
% Bibliography
% ------------------------------------------------------------------
\bibliographystyle{jfm}
\bibliography{bibl}

\end{document}